\newif{\ifarxiv}
\newif{\ifremarks}

\arxivtrue  % for stuff that should only go into the arxiv preprint, comment out for PRL version
\remarkstrue  % turn remarks on/off

\newcommand\arxor[2]{\ifarxiv{#1}\else{#2}\fi}

\ifremarks
\newcommand{\remarkab}[1]{{\renewcommand{\bfdefault}{b}\color[RGB]{0,150,150}{\textbf{A:~#1}}}}
\newcommand{\remarktb}[1]{{\renewcommand{\bfdefault}{b}\color[RGB]{0,150,0}{\textbf{T:~#1}}}}
\fi
\providecommand{\remarkab}[1]{\ignorespaces}
\providecommand{\remarktb}[1]{\ignorespaces}

\documentclass[
 reprint,
 amsmath,
 amssymb,
 aps,
 prl,
 nofootinbib,
 nobibnotes,
 longbibliography,
 preprintnumbers
]{revtex4-2}


\usepackage[a4paper,vmargin=2cm,hmargin=1.5cm]{geometry}

\usepackage[utf8]{inputenc}
\usepackage{comment}
\usepackage[pdftex]{graphicx}
\usepackage{wasysym}
\usepackage{amsfonts}
\usepackage{ifsym}

\usepackage[dvipsnames]{xcolor}
\usepackage{color}
\usepackage{graphicx}
\usepackage{dcolumn}
\usepackage{bm}
\usepackage{hyperref}

\usepackage{makecell}
\usepackage{multirow}
\usepackage{listings}
\usepackage{tabularray}
\usepackage{bbold}
\UseTblrLibrary{booktabs}

\lstdefinelanguage{Sage}{
	morekeywords={for, in, if, and}
	}
\hypersetup{plainpages=false}
\hypersetup{pdfpagemode=UseOutlines}
\hypersetup{bookmarksnumbered=true}
\hypersetup{bookmarksopen=true}
\hypersetup{pdfstartview=FitH}
\hypersetup{colorlinks=true}
\hypersetup{citecolor=[rgb]{0 .4 0}}
\hypersetup{urlcolor=[rgb]{.4 0 0}}
\hypersetup{linkcolor=[rgb]{0 0 .5}}

\newcommand{\namedref}[2]{\hyperref[#2]{#1~\ref*{#2}}}
\newcommand{\secref}[1]{\namedref{Section}{#1}}
\newcommand{\appref}[1]{\namedref{Appendix}{#1}}
\newcommand{\tabref}[1]{\namedref{Table}{#1}}
\newcommand{\figref}[1]{\namedref{Figure}{#1}}

\allowdisplaybreaks
\ifarxiv
\makeatletter
\def\paragraph{%
  \@startsection
    {paragraph}%
    {4}%
    {\z@}%
    {1.5ex \@plus.5ex \@minus .2ex}%
    {-1em}%
    {\normalfont\normalsize\bfseries}%
}%
\makeatother
\fi

\def\etal.{et\penalty50\ al.}
\usepackage{xspace}

\newcommand*{\ie}{i.\,e.\@\xspace}

\makeatletter\newcommand*{\etc}{%
    \@ifnextchar{.}%
        {etc}%
        {etc.\@\xspace}%
}\makeatother

\def\clap#1{\hbox to 0pt{\hss#1\hss}}

\usepackage[mathbold,autobold,greekcaps,greeklower]{mathfixs}
\usepackage{siunitx}

\usepackage{empheq}
\newlength{\widefboxpadding}
\newcommand{\nn}{\nonumber}

\newcommand{\subrm}[1]{_{\text{#1}}}

\newcommand{\grp}[1]{\mathrm{#1}}

\newcommand{\grSO}{\grp{SO}}

\newcommand{\filename}[1]{\texttt{#1}\xspace}
\newcommand{\software}[1]{\textsc{#1}\xspace}
\newcommand{\mathematica}{\software{Mathematica}}
\newcommand{\sage}{\software{Sage}}

\RequirePackage[extdef]{delimset}
\makeatletter
\providecommand{\brkleft}[1][r]{\begingroup\def\dlm@use{\delim(.}%
\if r#1 \def\dlm@use{\delim(.}\fi%
\if s#1 \def\dlm@use{\delim[.}\fi%
\if c#1 \def\dlm@use{\delim\{.}\fi%
\if a#1 \def\dlm@use{\delim<.}\fi%
\expandafter\endgroup\dlm@use}
\providecommand{\brkright}[1][r]{\begingroup\def\dlm@use{\delim.)}%
\if r#1 \def\dlm@use{\delim.)}\fi%
\if s#1 \def\dlm@use{\delim.]}\fi%
\if c#1 \def\dlm@use{\delim.\}}\fi%
\if a#1 \def\dlm@use{\delim.>}\fi%
\expandafter\endgroup\dlm@use}
\makeatother

\DeclareMathOperator{\tr}{Tr}

\newcommand{\op}[1]{\mathcal{#1}}
\newcommand{\order}{\mathcal{O}\brk}

\newcommand{\superN}{\mathcal{N}}
\newcommand{\Nc}{N\subrm{c}}

\newcommand{\dd}[2][]{\mathinner{d\ifx#1\empty\else{^#1}\fi#2}}

\newcommand{\lagr}{\mathcal{L}}
\newcommand{\Lint}{L\subrm{int}}

\newcommand{\x}[1]{x_{#1}^2}
\newcommand{\twentyprime}{\mathbold{20'}}
\begin{document}

\preprint{DESY-25-173}

\title{The Non-Planar Four-Point Integrand and Konishi Dimension \texorpdfstring{\\}{} in
\texorpdfstring{$\superN=4$}{N=4} Super Yang--Mills Theory at Five Loops}

\author{Till Bargheer}
\email{till.bargheer@desy.de}
\affiliation{Deutsches Elektronen-Synchrotron DESY, Notkestr.~85, 22607 Hamburg, Germany}
\author{Albert Bekov}
\email{albert.bekov@desy.de}
\affiliation{Deutsches Elektronen-Synchrotron DESY, Notkestr.~85, 22607 Hamburg, Germany}

\begin{abstract}
We compute the complete non-planar integrand for the correlation
function of four lightest scalar operators in $\superN=4$ super
Yang--Mills theory at five-loop order. This is equivalent to the
super-correlator of nine stress-tensor multiplets in the self-dual
theory. Starting with an ansatz of \mbox{$f$-graphs}, we impose constraints
from light-cone limits, and fix the remaining freedom by using the reformulation of the theory in twistor space.
We develop an efficient GPU-based algorithm for the numerical evaluation of the twistor rules.
Our result for the integrand suggests a sharpened \mbox{$f$-graph} genus conjecture.
As an application, we extract the
five-loop non-planar anomalous dimension of the Konishi operator,
whose $1/\Nc^2$ term is entirely given in terms of odd zeta values of
transcendentality five and higher. Our
code and result are provided in ancillary files.
\end{abstract}

% Abstract proposed by Google Gemini:

% We determine the complete five-loop integrand of the four-point
% correlation function of stress-tensor multiplet operators in
% super-Yang-Mills theory, including all non-planar corrections. This
% result is made possible by a novel high-performance computational
% framework for the exact rational reconstruction of Grassmann
% structures, which overcomes the long-standing bottleneck of high-order
% non-planar contractions. Beyond providing the first full determination
% of the five-loop non-planar Konishi anomalous dimension, our results
% reveal a surprising structural property: the integrand is composed
% exclusively of f-graphs whose genus exactly matches the expansion
% order. This discovery leads us to propose a sharpened genus-expansion
% conjecture that constrains the theory’s analytic structure across all
% orders. Our findings provide essential benchmarks for integrability
% and the AdS/CFT correspondence, while the introduced GPU-accelerated
% methodology opens a new path for precision calculations in non-planar
% gauge theories.

\maketitle

%%%%%%%%%%%%%%%%%%%%%%%%%%%%%%%%%%%%%%%%%%%%%%%%%%%%%%%%%%%%
%%%%%%%%%%%%%%%%%%%%%%%%%%%%%%%%%%%%%%%%%%%%%%%%%%%%%%%%%%%%
\section{Introduction}
\label{sec:introduction}

The correlator of four lightest ($\twentyprime$) scalar BPS operators
$
\mathcal{O}(x, y) = \tr\brk[s]{(y \cdot \phi(x))^2}
$
in $\superN=4$ SYM has been studied extensively at
weak~\cite{Gonzalez-Rey:1998wyj,
Eden:1998hh, Eden:1999kh, Eden:2000mv, Bianchi:2000hn, Eden:2011we, Eden:2012tu,
Ambrosio:2013pba, Bourjaily:2015bpz, Bourjaily:2016evz, He:2024cej,
Bourjaily:2025iad}
as well as strong and finite coupling~\cite{DHoker:1999pj,
DHoker:1999ni, Arutyunov:2000py, Rastelli:2016nze, Alday:2017xua,
Aprile:2017bgs, Rastelli:2017udc, Yuan:2018qva, Carmi:2019ocp,
Drummond:2019hel, Huang:2021xws, Drummond:2022dxw, Caron-Huot:2022sdy,
Huang:2024dck, Caron-Huot:2024tzr}.
It is the simplest correlator with a non-trivial coupling dependence,
contains information on many observables through OPE and cusp
limits, and is related to phenomenologically interesting quantities
such as event shapes and energy-energy
correlators~\cite{Belitsky:2013xxa,Belitsky:2013bja,Belitsky:2013ofa}.

At weak coupling, the loop correlator can be computed by
integrating
over Lagrangian insertions~\cite{Intriligator:1998ig,
Eden:1999kw, Eden:2000mv},
\begin{align}
	&G_4 \equiv \langle \mathcal{O}(x_1, y_1)\, \mathcal{O}(x_2, y_2)\, \mathcal{O}(x_3, y_3)\, \mathcal{O}(x_4, y_4)\, \rangle \nonumber\\
	&\quad\; = \sum_{\ell \ge 0} a^\ell  \int \frac{d^4x_5\ldots d^4x_{4+\ell}}{(-1)^\ell\, \ell !} \, G_4^{(\ell)}(x_i, y_i)\,,
    \\ &
    G_4^{(\ell)}(x_i, y_i) =
    \brk[a]{\op{O}_1\dots\op{O}_4\,\lagr_5\dots\lagr_{4+\ell}}_0
    \,,
\end{align}
where $\lagr_i=\lagr(x_i)$ is the chiral interaction Lagrangian,
$\brk[a]{\dots}_0$ is the leading-order correlator, and
$a = g_\mathrm{YM}^2 \Nc/(4 \pi^2)$ is the 't~Hooft coupling.
Since $\op{O}$ and $\lagr$ are components of the chiral
stress-tensor multiplet,
\begin{equation}
	\mathcal{O}(x, y, \theta) = \mathcal{O}(x, y) + \ldots + (\theta)^4 \mathcal{L}(x)\,,
\end{equation}
the \emph{integrand} $G_4^{(\ell)}$ inherits an $\grp{S_{4+\ell}}$
permutation symmetry from the supercorrelator after factoring out a
universal superinvariant, which allowed to
bootstrap the planar integrand to twelve
loops~\cite{Eden:2011we,Eden:2012tu,Ambrosio:2013pba, Bourjaily:2015bpz, Bourjaily:2016evz, He:2024cej, Bourjaily:2025iad}.
The non-planar integrand is known to four
loops~\cite{Eden:2012tu,Fleury:2019ydf}. Here, we push the computation of the
non-planar integrand to five loops.

We explain our method in \secref{sec:method}, discuss the result
in \secref{sec:result}, and extract the non-planar five-loop
Konishi anomalous dimension in \secref{sec:konishi}.
Technical details are presented in \arxor{appendices}{the
Supplemental Material}. Our code and result are available in ancillary
files.

%%%%%%%%%%%%%%%%%%%%%%%%%%%%%%%%%%%%%%%%%%%%%%%%%%%%%%%%%%%%
%%%%%%%%%%%%%%%%%%%%%%%%%%%%%%%%%%%%%%%%%%%%%%%%%%%%%%%%%%%%
\section{Method}
\label{sec:method}

We adopt the approach used in the four-loop computation
\cite{Fleury:2019ydf}, utilizing the twistor reformulation of the theory. Due to its
larger gauge freedom, this framework offers a simpler set of Feynman
rules, at the cost of introducing a fixed reference twistor, which does
not affect the final result, yet is practically impossible to eliminate.
We therefore resort to using a manifestly invariant ansatz for
the non-planar integrand, and fix its coefficients by numerically
matching against the twistor answer. This is particularly efficient for
the four-point integrand, where a compact ansatz in so-called $f$-graphs is available.

The main difficulty at five loops,
beyond the overall large number of graphs, are the contractions of 20 Grassmann
variables introduced in the twistor Feynman rules. These make
the numerical evaluation very costly, such that it cannot be performed
on an average computer. We solve this problem by implementing a fast GPU-based algorithm, which substantially accelerates the contraction process.

%%%%%%%%%%%%%%%%%%%%%%%%%%%%%%
\paragraph*{$f$-graph ansatz.}

The Born-level correlator of four stress-tensor operators and $\ell>0$ chiral on-shell
Lagrangians takes the form \cite{Eden:2000bk, Eden:2012tu}
\begin{equation} \label{eq:integrand}
	G_4^{(\ell)}(x_i, y_i) = 2\frac{\Nc^2-1}{(4\pi^2)^{4+\ell}}\,R(x_i, y_i)\,\xi_4\,F^{(4+\ell)}(x_i)\,,
\end{equation}
where $R(x_i,y_i)$ is the unique superconformal invariant of four points and
$\xi_4=\x{12}\x{13}\x{14}\x{23}\x{24}\x{34}$. $F^{(n)}$ is a
rational function of $x_{ij}^2$ that is invariant under all permutations of
$x_1,\dots,x_n$, has at most simple poles in $\x{ij}$, and conformal
weight $+4$ in each point $x_i$. This function can be expanded in
so-called $f$-graphs,
$F^{(n)}=\sum_{i=1}^{\mathcal{N}_n}c_i^{(n)}f^{(n)}_i$, where each
$f^{(n)}_i$ represents a permutation-invariant sum of terms,
normalized such that each term appears with unit coefficient. The total number of
$f$-graphs $\mathcal{N}_n$ is listed in \tabref{tab:fgraphs}.
The color dependence of $G_4^{(\ell)}$ is encoded in the coefficients
$c_i^{(n)}$, which are polynomials in $1/\Nc^2$. Thus we can define the genus expansion of $F^{(n)}$ by
\begin{equation}
	F^{(n)} = \sum_{g\ge 0} \frac{1}{\Nc^{2g}}F^{(g,n)} = \sum_{g\ge 0} \frac{1}{\Nc^{2g}}\sum_{i=1}^{\mathcal{N}_n}c_i^{(g,n)}f^{(n)}_i\,,
    \label{eq:ansatz}
\end{equation}
where $c_i^{(g,n)}$ are now purely numerical coefficients.

It has been conjectured
that the integrand $G_4^{(g, \ell\ge2)}$ only depends on $f$-graphs up to genus $g$~\cite{Eden:2012tu}:
\begin{equation} \label{eq:fGraphCond}
	c_i^{(g,n)} = 0 \quad\text{if}\quad \mathrm{genus}(f^{(n)}_i)>g \,,
\end{equation}
where the genus of an $f$-graphs is defined as the minimal genus of
the simple, connected graph obtained by associating an edge to each
squared distance in the denominator.
For planar integrands ($g=0$), this implies that only the planar $f$-graphs contribute, which has been verified up to twelve loops~\cite{Bourjaily:2025iad}.
In practice, we find the genus by using the algorithm
\texttt{multi\_genus} of~\cite{Brinkmann:2022},
see \tabref{tab:fgraphs} for the statistics.

In general, not all $f$-graphs are linearly independent.
Since they are functions of squared distances $\x{ij}$, they satisfy conformal
Gram identities.
These emerge when considering the $(n\times n)$ Gram matrix
with elements
\begin{equation} \label{eq:gram}
	\mathcal{G}_{ij} = X_i \cdot X_j = \x{ij} \quad\text{for}\quad i, j = 1,\ldots, n \,,
\end{equation}
where $X_i$ are the six-dimensional null vectors on which the
conformal symmetry $\grSO(2,4)$ acts linearly. This matrix is of rank
six, implying that all $(7\times 7)$-minors vanish. After dividing by
appropriate monomials in $\x{ij}$, such that each point is of
conformal weight $+4$, and symmetrizing over all $n$ points, these
conditions generate the Gram identities between the $f$-graphs.

\begin{table}[tb]
	\centering

    \begin{tblr}{colspec={
            Q[c,m,wd=9mm]
            Q[c,m,wd=9mm]
            Q[c,m,gray9,wd=9mm,valign=b]
            Q[c,m,gray9,colsep=0pt,wd=9mm,valign=b]
            Q[c,m,gray9,wd=9mm,valign=b]
            Q[c,m,wd=9mm,valign=b]
        }}
        \toprule
        % &&& genus &&\\[-0.7ex]
        % \cline[0.1pt]{3,4,5}
        $n$ & $\mathcal{N}_n$ & {$0$} & {genus\\[0.8ex]$1$} & {$2$} & Gram id. \\
        \midrule
        7  & 4 & 1 & 3 & -- & 1\\
        8  & 32 & 3 & 29 & --  & 3\\
        9 & 930 & 7 & 833 & 90 & 208\\
        \bottomrule
	\end{tblr}
	\caption{Numbers of $f$-graphs $\mathcal{N}_n$ entering the
    three-, four-, and five-loop ansatz, their genus distribution, and
    the number of linearly independent Gram identities.}
	\label{tab:fgraphs}
\end{table}

Determining the four-point integrand amounts to finding the numerical
coefficients $c^{(g,n)}_k$ in the ansatz~\eqref{eq:ansatz}.
The freedom in the ansatz can be greatly reduced by imposing null or
coincidence limits. At the planar level, such limits in fact uniquely
determine the integrand up to twelve loops \cite{Eden:2011we,
Eden:2012tu, Ambrosio:2013pba, Bourjaily:2015bpz, Bourjaily:2016evz,
He:2024cej, Bourjaily:2025iad}.
For the subleading $1/\Nc^2$ terms, this is no longer the case: Some
free coefficients remain, which have to be fixed
by a numerical match with the twistor computation.

In order to constrain $F^{(g\ge1, 9)}$, we impose the light-cone
constraint~\cite{Eden:2012tu}: Setting neighboring external operators
to null separation ($\x{12},\x{23},\x{34},\x{41}\rightarrow 0$), the
tree-level normalized correlator takes the form
\begin{equation} \label{eq:llcorrelator}
	\frac{G_4}{G_4^{(0)}} = 1 + 2 \sum_{\ell\ge1}\frac{a^\ell}{(-4\pi^2)^\ell}\, \int \left(\prod_{k=5}^{4+\ell}d^4x_k\right)\, \mathcal{F}^{(4+\ell)}(x_i)\,,
\end{equation}
where $\mathcal{F}^{(n)}=\lim_{\x{i,i+1}\rightarrow0}\x{13}\x{24}\xi_4
F^{(n)}/\ell!$ is the null-square limit of the tree-level normalized
sum of $f$-graphs. While $\mathcal{F}^{(n)}$ is finite, the integrals
in~\eqref{eq:llcorrelator} diverge: Starting from $\ell = 2$, the $\ell$-loop
contribution to the logarithm of~\eqref{eq:llcorrelator} exhibits a
$\log^{\ell}(x_{i,i+1}^2)$
divergence~\cite{Alday:2010zy}. However, the integrals of individual
$f$-graphs generally produce stronger
divergences $\log^k(x_{i,i+1}^2)$, $k>\ell$. These divergences arise from integration
regions where one of the internal operators approaches one of the
light-like segments. To ensure the correct
logarithmic divergence of the correlator, we require the
\emph{light-cone constraints}:%
\footnote{For the planar integrand, given by an ansatz in terms of planar $f$-graphs, this constraint fixes all seven coefficients.}
In the
limit $x_5 \rightarrow \alpha x_1 + (1 - \alpha)x_2$, the integrand of
the logarithm of~\eqref{eq:llcorrelator} must vanish order by order in
$\alpha$. For the non-planar five-loop integrand, these conditions are:%
\footnote{There are no further terms in~\eqref{eq:lc-constraint1}
because $\mathcal{F}^{(1,n)}=0$ for $n<8$.}
\begin{align}
    \label{eq:lc-constraint1}
	&\lim_{x_5\rightarrow \alpha x_1 + (1-\alpha)x_2} \x{15}\x{25} \big(\mathcal{F}^{(1, 9)} -2 \mathcal{F}^{(1, 8)}\mathcal{F}^{(0, 5)}\big) = 0
    \,, \\
    \label{eq:lc-constraint2}
	&\lim_{x_5\rightarrow \alpha x_1 + (1-\alpha)x_2} \x{15}\x{25} \,\mathcal{F}^{(g\ge2,\, 9)} = 0
    \,.
\end{align}
For the genus-one integrand, following~\eqref{eq:fGraphCond}, one can impose that the 90
genus-two $f$-graphs do not contribute to the final integrand.
Then, imposing~\eqref{eq:lc-constraint1} fixes all but 155
coefficients of $F^{(1,9)}$, of which 134 parametrize Gram identities.
Hence, only $21$ coefficients are left to be determined by the twistor computation.%
\footnote{We further tested the five-loop ansatz against the
cusp-limit constraint of~\cite{He:2024cej}. This constraint turns out
to be less restrictive and is already fully captured by the light-cone
constraints.}
Starting from genus two, all 930 $f$-graphs contribute,
and~\eqref{eq:lc-constraint2} fixes all but 23 coefficients,
accompanied by 208 Gram identities, at every order in the genus
expansion.

%%%%%%%%%%%%%%%%%%%%%%%%%%%%%%
\paragraph*{Twistor computation.}

We briefly summarize the steps necessary to compute the integrand of
the four-point correlator using the twistor formulation. More details
can be found in \arxor{\appref{sec:twistors}}{the Supplemental Material, Appendix B}.

First, all relevant skeleton graphs are generated. At $\ell$-loop
order, they are specified as all connected $(4+\ell)$-point
graphs with $4+2\ell$ edges and a valency of at least~2 at each
vertex, and can for example be obtained with the open-source program
\sage~\cite{sagemath} (see \tabref{tab:twistor}).

Second, starting from each skeleton graph $\mathfrak{g}$, all
associated ribbon graphs $\gamma\in\Gamma_\mathfrak{g}$ can be formed by
permuting over the order of the edges around each vertex. Then, the
twistor Feynman rules are applied to these ribbon graphs: Each
graph is dressed with a factor $(g_\mathrm{YM}^2)^{P-V}(4\pi^2)^{-P}$,
with $P$ and $V$ denoting the number of edges (propagators) and
vertices of the graph.
Every edge connecting vertices $i$ and $j$ is
dressed with a propagator $d_{ij}=y_{ij}^2/\x{ij}$. Each vertex $i$
connected to cyclically ordered vertices $j_1,\ldots,j_k$ produces an $R$-factor
$R^i_{j_1\ldots j_k}$ and a color trace $\tr(T^{a_{ij_1}}\ldots
T^{a_{ij_k}})$, where $T^a$ denote the $\grp{SU}(\Nc)$ generators in
the fundamental
representation.
For each ribbon graph $\gamma$, the traces can be evaluated to a function in $\Nc$ by successively applying the fusion and fission rules
\begin{align}
	\tr(T^a B T^a C) &= \tr(B)\tr(C)-\tr(BC)/\Nc \,,\\
	\tr(BT^a)\tr(CT^a) &= \tr(BC)-\tr(B)\tr(C)/\Nc\,.
\end{align}
This yields an overall factor of $(\Nc^2-1)\Nc^\ell$
multiplied by a polynomial $c_\gamma(1/\Nc^2)$.
Furthermore, the dependence of the $R$-factors on the ribbon structure
can be canonicalized by pulling out factors $\langle ijk\rangle$
via \arxor{(cf.~\eqref{eq:onshellcond})}{}
\begin{equation}
	R^i_{j_{\tau(1)}\ldots j_{\tau(k)}} = R^i_{j_{1}\ldots j_{k}}\frac{\langle ij_1 j_2 \rangle\ldots\langle ij_k j_1 \rangle}{\langle ij_{\tau(1)}j_{\tau(2)} \rangle\ldots\langle ij_{\tau(k)}j_{\tau(1)} \rangle} \,,
\end{equation}
where $\tau$ is a permutation.
Let us denote the product of these kinematic factors for each
canonicalized ribbon graph by $s_\gamma(\langle ijk \rangle)$. In this
way, a universal product of propagators and $R$-factors can be pulled
out of the sum over all ribbon graphs $\Gamma_\mathfrak{g}$ associated
to the same skeleton graph $\mathfrak{g}$, leaving a sum that
entails the complete dependency on
the color structure $c_\gamma(1/\Nc^2)$, and includes some kinematic
factors $s_\gamma(\langle ijk \rangle)$. Splitting the $R$-factors
into basic factors $R_{jkl}^i$ of Grassmann degree two \arxor{(cf.~\eqref{eq:Rsplit})}, the sum of ribbon graphs over
$\Gamma_\mathfrak{g}$ can be schematically written as
\begin{align}
	&\sum_{\gamma \in \Gamma_\mathfrak{g}} \gamma = a^\ell\,\frac{\Nc^2-1}{(4\pi^2)^{4+\ell}}\,p_\mathfrak{g}(1/\Nc^2, \langle ijk \rangle)\prod^{4+2\ell}_\mathfrak{g} d_{ij} \times \prod^{2\ell}_\mathfrak{g} R^i_{jkl} \nonumber\\
	&\text{with}\quad p_\mathfrak{g}(1/\Nc^2, \langle ijk \rangle) =  \sum_{\gamma \in \Gamma_\mathfrak{g}} c_\gamma(1/\Nc^2) \,s_\gamma(\langle ijk \rangle)\,.
\end{align}
Here, the symbol $\mathfrak{g}$ below the product signs indicate that
the products only depend on the skeleton graph, not on the specific
ribbon graph, while the number above denotes the number of factors in
each product. The prefactors $p_\mathfrak{g}$ encode the genus
expansion, and can be evaluated for each skeleton graph at this stage.
It is known that the full integrand is
completely planar up to three loops, and
there is a single correction of order
$\mathcal{O}(1/\Nc^2)$ at four loops. We find that the same is true at
five loops: The prefactors $p_\mathfrak{g}$ exhibit at most a $\order{1/\Nc^2}$
contribution. This implies that
$F^{(g\ge2,\,9)}$
vanishes, leaving only the remaining 21 coefficients of $F^{(1, 9)}$ to be determined.

Third and finally, each skeleton graph must be permuted over all
inequivalent vertex labelings. This yields a large number of graphs $\mathfrak{G}$, in terms of which the integrand~\eqref{eq:integrand} can now be schematically expressed as
\begin{equation}
	G_4^{(\ell)} = \frac{\Nc^2-1}{(4\pi^2)^{4+\ell}} \sum_{\mathfrak{g}\in\mathfrak{G}}\,\left(p_\mathfrak{g}\,\prod^{4+2\ell}_\mathfrak{g} d_{ij} \times \prod^{2\ell}_\mathfrak{g} R^i_{jkl}\right)\,.
    \label{eq:twistorFinal}
\end{equation}
To reduce the computational time of evaluating the expression, we aim
to reduce the number of permuted graphs $\mathfrak{G}$ to consider. As
in~\cite{Fleury:2019ydf}, we choose a polarization in which all
$y_{ij}^2$ vanish except $y_{12}^2$ and $y_{34}^2$, which drastically
cuts down the number of contributing graphs.
Yet, the complete integrand is still accessible,
since the polarizations only appear in the four-point invariant
$R(x_i,y_i)$ in~\eqref{eq:integrand}. Moreover, we can eliminate
graphs that cannot generate the desired Grassmann monomial
$\theta_5^4\dots\theta_{4+\ell}^4$. For example, $R$-factors only
depending on the external operators can be set to zero, and
products of $R$-factors in which any internal label appears
fewer than twice can also be removed. Listing the permutations,
specifying the polarization, and removing non-contributing
products of $R$-factors, we arrive at the second row of
\tabref{tab:twistor}.
Some remaining graphs still evaluate to zero, despite it not being apparent from
their symbolic expression. We remove such graphs by probing all remaining products of $R$-factors numerically, and
keeping only graphs that are indeed non-vanishing. Their counts appear in the third row of \tabref{tab:twistor}.

\begin{table}[tb]
	\centering
    \begin{tblr}{colspec={
            Q[l,m,colsep=4pt]
            Q[r,m,colsep=4pt]
            Q[r,m,colsep=4pt]
            Q[r,m,colsep=4pt]
            Q[r,m,colsep=4pt]
            Q[r,m,colsep=4pt]
        }}
        \toprule
        $\mspace{80mu}\ell$ & 1 & 2 & 3 & 4 & 5 \\
        \midrule
        \# skeleton graphs & 3 & 11 & 63 & 513 & 5,553 \\
        \# permuted graphs & 1 & 73 & 6,321 & 732,288 & 110,732,441 \\
        \# non-zero graphs & 1 & 73 & 5,901 & 627,148 & 87,757,390\\
        \bottomrule
	\end{tblr}

	\caption{First row: Number of skeleton graphs generated with
    \sage. Second row: Number of permuted and polarized graphs, excluding graphs
    that cannot produce the correct Grassmann monomial. Third row:
    Final number of graphs that have a non-vanishing
    contributions to the twistor result.}
	\label{tab:twistor}
\end{table}

Let us mention that in four dimensions, the integrand $G_4^{(\ell)}$ has a
pseudo-scalar contribution, which is produced by a total derivative
term in the twistor-reformulated action of $\mathcal{N}=4$, and
therefore vanishes upon integration. In Lorentzian signature,
this contribution is imaginary, and hence can be numerically isolated.
However, we perform the calculation in split signature, where the
twistor computation remains real, such that we can efficiently employ finite fields.
In order to isolate the pseudo-scalar contribution, we compute each numerical
point twice, with kinematics that differ by a parity transformation. Taking their
mean removes any pseudo-scalar contribution.

%%%%%%%%%%%%%%%%%%%%%%%%%%%%%%
\paragraph*{Grassmann contraction.}

One of the main challenges in evaluating the twistor expression
numerically lies in the contraction of the
Grassmann numbers that appear in the $R$-factors.
Each term in the five-loop twistor
expression~\eqref{eq:twistorFinal} contains ten
$R$-factors $\{R_1,\ldots,R_{10}\}$, each quadratic in the Grassmann
variables $\theta^{a\alpha}$, that must be contracted into the unique
product $\theta_5^4\dots\theta_{9}^4$ involving all 20
Grassmann numbers.
Splitting each $R_a$ into a list of the $\binom{20}{2}=190$ products of
Grassmann numbers,
we find that the most
efficient implementation iteratively
contracts the $R$-factors into successively higher-degree products of
Grassmann numbers as follows:
\begin{align} \label{eq:grassmann}
	(S_{a})^k = (t_{a-1})^k_{ij} (S_{a-1})^i (R_a)^j \quad\text{with}\quad a \in [2,\,10]\,,
\end{align}
with $S_1 = R_1$, and $t_a$ being fixed numerical sparse tensors that contract the
vectors at each intermediate step. These successive
tensor–vector–vector multiplications can be implemented highly
efficiently on a graphics processing unit (GPU). Details on the
tensors $t_a$ and the GPU-based implementation of this algorithm are
provided in~\arxor{\appref{app:grassmann}}{the Supplemental Material, Appendix C}. Under optimal settings, the
Grassmann contractions required to evaluate~\eqref{eq:twistorFinal} at
a single numerical point take approximately 17 hours on a single
\textit{Nvidia A100} GPU, rendering the determination of the 21
coefficients (each requiring the evaluation at two numerical points)
in the five-loop genus-one ansatz practically feasible.

%%%%%%%%%%%%%%%%%%%%%%%%%%%%%%%%%%%%%%%%%%%%%%%%%%%%%%%%%%%%
%%%%%%%%%%%%%%%%%%%%%%%%%%%%%%%%%%%%%%%%%%%%%%%%%%%%%%%%%%%%
\section{Result}
\label{sec:result}

As discussed above, the full five-loop integrand only consist of the
planar part and the genus-one correction. By numerically
matching against the twistor result, we can fix the remaining
$21$ free parameters in the genus-one ansatz, and find the full five-loop
integrand.
Of the $930$ nine-point $f$-graphs, we excluded the $90$ genus-two
graphs, as they do not contribute to the genus-one integrand. By
adding an appropriate combination of the $134$ Gram identities among
the remaining graphs, we obtain a particular solution for the
remaining $840$ $f$-graph coefficients that exhibits the following
properties: First, $582$ coefficients are zero, thus the corresponding
$f$-graphs do not contribute to
the genus-one integrand.
Second, the remaining $258$ non-zero coefficients are of small integer
value, as is visualized in~\arxor{figref{fig:histogram}
in~\appref{app:grassmann}}{the Supplemental Material, Appendix C, Figure $1$}. We provide this solution in the ancillary
file \filename{solution.m}.%

Furthermore, we notice that a stronger statement
than~\eqref{eq:fGraphCond} might hold, namely the following:
\textit{The integrand $\mathit{G_4^{(g,\ell\ge2)}}$ only depends on
$\mathit{f}$-graphs of exactly genus~$\mathit{g}$}, that is
\begin{equation}
	c_i^{(g,n)} = 0 \quad\text{if}\quad \mathrm{genus}(f^{(n)}_i)\neq g \,.
\end{equation}
First, this is consistent with the non-planar integrands found so far.
The genus-one correction to the four-loop integrand~\cite{Eden:2012tu,Fleury:2019ydf} can be uniquely
written in terms of the $29$ genus-one eight-point $f$-graphs, using the three
Gram identities to remove the dependency on the planar graphs. At five
loops, we can construct a result in terms of only the
genus-one $f$-graphs, reducing the Gram freedom to~$127$. Second, it
immediately rules out $g\ge2$ and $g\ge3$ correction to the four- and
five-loop integrand, respectively. Regarding the genus-two correction to
the five-loop integrand, this statement implies that it is given in
terms of the $90$ genus-two nine-point $f$-graphs. Imposing the light-cone
constraint~\eqref{eq:lc-constraint2} on this ansatz sets all but one
coefficient to zero. The remaining $f$-graph vanishes in this
limit, and thus trivially satisfies the constraint.%
\footnote{An explicit expression for this $f$-graph is provided in
line 843 of the ancillary file \filename{allFGraphs.txt}. This
behavior was already observed for the ``8-point version'' of this
$f$-graph (cf.~(5.25) in~\cite{Eden:2012tu}).}
%

%%%%%%%%%%%%%%%%%%%%%%%%%%%%%%%%%%%%%%%%%%%%%%%%%%%%%%%%%%%%
%%%%%%%%%%%%%%%%%%%%%%%%%%%%%%%%%%%%%%%%%%%%%%%%%%%%%%%%%%%%
\section{Konishi anomalous dimension}
\label{sec:konishi}

One of the immediate observables obtainable from the four-point
$\twentyprime$ integrand is the anomalous dimension $\gamma_{\mathcal{K}}(a)$ of the Konishi
operator $\mathcal{K}(x) = \tr(\phi(x) \cdot \phi(x))$, which is
encoded in its two-point function. The latter can be extracted from
the four-point correlator via the operator product expansion (OPE).
$\mathcal{K}$ dominates the OPE, thus $\gamma_{\mathcal{K}}(a)$ can be
extracted by by double-pinching the integrand ($x_2\rightarrow x_1$ and
$x_4\rightarrow x_3$)~\cite{Eden:2012fe}:
\begin{multline} \label{eq:dpcorrelator}
	\log\brk4{1+6\,\sum_{\ell\ge1} \frac{a^\ell}{(-4\pi^2)^\ell}\, \int \left(\prod_{k=5}^{4+\ell}d^4x_k\!\right)\, \hat{F}^{(4+\ell)}(x_i)}
	\\
	= \frac{1}{2}\gamma_\mathcal{K}(a) \log(u) + \mathcal{O}(u^0)
	\,,
\end{multline}
with $u=\x{12}\x{34}/(\x{13}\x{24})$ being one of the conformal
cross-ratios and
\begin{equation} \label{eq:OPEfgraphs}
\hat{F}^{(n)} = \lim_{\substack{x_2\rightarrow
x_1\\x_4\rightarrow x_3}}x_{13}^4\xi_4 F^{(n)}/\ell!
\,.
\end{equation}
Expanding the
above equation to the genus-one five-loop order yields
\begin{equation} \label{eq:konishiIntegral}
	\int \brk*{\prod_{k=5}^{9}d^4x_k\!} I^{(1,5)}
    = \frac{1}{2}\gamma_\mathcal{K}^{(1,5)} \log(u) + \mathcal{O}(u^0)\,,
\end{equation}
with $\gamma_\mathcal{K}^{(1,5)}$ denoting the genus-one five-loop correction to
the Konishi anomalous dimension, and the integrand given by $I^{(1,5)}
= -6(\hat{F}^{(1,5)}-6\hat{F}^{(0,1)}\hat{F}^{(1,4)})/(4\pi^2)^5$. The
log-divergence of the above integral arises when all of the
integration points approach one of the external points $x_1$ or $x_3$.
In order to extract the coefficient $\gamma_\mathcal{K}^{(1,5)}$ in
front of this divergence, we follow the steps outlined
in~\cite{Eden:2012fe}: First, we take $x_3\rightarrow\infty$,
effectively replacing all $\x{3k}\rightarrow\x{13}$, to single out
the divergence that arises when all points approach $x_1$, and define
$\hat{I}^{(1,5)}=\lim_{x_3\rightarrow \infty}I^{(1,5)}$. Second, we
apply dimensional regularization with $d=4-2\epsilon$ and perform all
but one integration
\begin{equation} \label{eq:konishiInt}
	2 \int\brk*{\prod_{k=6}^{9}d^{4-2\epsilon}x_k\!} \hat I^{(1,5)}
    = \frac{C^{(1,5)}}{\pi^2} (\x{15})^{-2-4\epsilon}\,,
\end{equation}
where $C^{(1,5)}$ is finite for $\epsilon\rightarrow0$. The factor 2
is included to take into account the contribution of the integration
around $x_3$. Third, comparing with~\eqref{eq:dpcorrelator}, a careful
analysis on the infrared rearrangement of the remaining integration
allows to identify the log-divergence by
\begin{equation} \label{eq:logdiv}
-2 \int \frac{d^{4-2\epsilon} x_5}{\pi^2} (\x{15})^{-2-4\epsilon}
=\log u
+ \order{u^0}
+ \order{\epsilon}\,,
\end{equation}
in the leading behavior in $\epsilon$. We refer for details on this
regularization scheme to App.~A in~\cite{Eden:2012fe}. Finally, we
identify $\gamma_\mathcal{K}^{(1,5)} =-C^{(1,5)}$,
obtaining the correction to the anomalous dimension.

Practically, obtaining the genus-one five-loop anomalous dimension
boils down to performing the four-loop propagator-type integration
in~\eqref{eq:konishiInt}. Starting from the obtained genus-one
five-loop integrand $G^{(1,5)}_4$ and the known result of
$G^{(1,4)}_4$~\cite{Eden:2012tu,Fleury:2019ydf}, we construct $\hat I^{(1,5)}$ and
find that $C^{(1,5)}$ is a linear
combination of $2855$ two-point four-loop integrals.
Applying the
integration-by-parts method, we reduce this linear combination of
integrals to a small set of master integrals using the program \texttt{FIRE6}~\cite{Smirnov:2019qkx,Smirnov:2023yhb}.
In particular, we find that it consists of 24 master integrals, 20 of
which are associated to planar graphs, while the remaining integrals
are associated to non-planar ones:
\begin{equation} \label{eq:masterInt}
	\gamma_\mathcal{K}^{(1,5)} =
    \eval[s]*{\,
    \sum_{k=1}^{20} v_k\,\mathcal{J}_k
    + \sum_{k=1}^{4} w_k\,\mathcal{I}_k
    }_{\mathrlap{\epsilon\rightarrow0}}\,.
\end{equation}
The coefficient functions $v_k$ and $w_k$ are expressed as expansions
in $\epsilon$. In particular, the leading order of each $w_k$ is $-1$.
By introducing dual momenta, the master integrals corresponding to planar graphs
can be transformed to four-loop propagator-type integrals in
momentum representation. Graphically, these momentum master integrals are represented by the dual graphs of the original planar ones.
The latter were explicitly calculated as
expansions in $\epsilon$ in~\cite{Baikov:2010hf,Lee:2011jt}, carried
out to sufficiently high orders, such that~\eqref{eq:masterInt} can be
evaluated at $\mathcal{O}(\epsilon^0)$. Of the four remaining master
integrals associated to non-planar graphs, $\mathcal{I}_1$ and
$\mathcal{I}_2$ were derived in~\cite{Eden:2012fe} (cf.~App.~B
therein). These integrals already appeared in the computation of the
planar five-loop correction to the Konishi anomalous dimension. The
explicit form and derivation of the final two master integrals
$\mathcal{I}_3$ and $\mathcal{I}_4$ can be found in~\arxor{\appref{app:masterInt}}{the Supplemental Material, Appendix D}.

Gathering the results of all master integrals and plugging them
into~\eqref{eq:masterInt}, finally yields the non-planar contribution
to the anomalous dimension of the Konishi operator at five loops:
\begin{equation}
\fbox{$\displaystyle
\gamma_\mathcal{K}^{(1,5)} = 135 \zeta_5 -\frac{243}{4} \zeta_3^2 + \frac{11907}{32} \zeta_7
\approx
427.406
\,.
$}
\label{eq:KonishiDim}
\end{equation}
Importantly, all poles in $\epsilon$ vanish in the final result of the
anomalous dimension, which is a strong consistency check of the
derivation. Furthermore, $\gamma_\mathcal{K}^{(1,5)}$ does not depend
on even Zeta values, which agrees with general
expectations~\cite{Broadhurst:1995km,Brown:2012ia} as well as existing
data~\cite{Leurent:2013mr,Velizhanin:2009gv}.
More subtly, there are also no
terms with transcendentality weight zero and three. This is also the
case for the non-planar correction of the four-loop Konishi anomalous
dimension~\cite{Velizhanin:2009gv}.

%%%%%%%%%%%%%%%%%%%%%%%%%%%%%%%%%%%%%%%%%%%%%%%%%%%%%%%%%%%%
%%%%%%%%%%%%%%%%%%%%%%%%%%%%%%%%%%%%%%%%%%%%%%%%%%%%%%%%%%%%
\section{Conclusion}
\label{sec:conclusion}

We have computed the full non-planar five-loop integrand of the
correlator of four $\twentyprime$ operators. The non-planar part is
proportional to $1/\Nc^2$, and it can be written exclusively in terms of genus-one
$f$-graphs, with small integer coefficients.
There are no genus-two contributions at
five loops, which is consistent with all existing data.
We also extracted the
non-planar Konishi anomalous dimension~\eqref{eq:KonishiDim}, which
consists of $\zeta_5$, $\zeta_3^2$, and $\zeta_7$ terms.
The dependence on $\Nc$ signifies that the
radius of convergence of the perturbative series shrinks at finite
$\Nc$ as expected, see~\arxor{\appref{sec:konishiFull}}{the Supplemental Material, Appendix A}.

We obtained the integrand from a combination of physical (light-cone)
constraints, and input from a numerical twistor computation.
An open question is whether the non-planar integrand could instead be fully determined by analytic constraints alone.
For example, the
non-planar part of the logarithm of $G_4/G_0$ may have at most
$\log^{\ell-1}$ singularities at~$\ell$ loops in the light-cone
limit~\cite{Eden:2012tu}, which is not necessarily fully imposed
by~\eqref{eq:lc-constraint1}, \eqref{eq:lc-constraint2}.
Further constraints could be derived by matching the $f$-graph periods~\cite{Zhang:2024ypu}
against the known completely integrated correlator~\cite{Dorigoni:2021bvj,Dorigoni:2021guq}.

Our result encodes a wealth of five-loop non-planar OPE data,
including the Konishi OPE coefficient~\cite{Georgoudis:2017meq}, the
twist-two spin-$J$ anomalous dimensions and OPE
coefficients~\cite{Velizhanin:2009gv,Eden:2012fe,Fleury:2019ydf}, and
the cusp anomalous dimension~\cite{Henn:2019swt}.
Extracting some of this data would provide valuable insight,
particularly also because of their relevance to the respective QCD counterparts.

%%%%%%%%%%%%%%%%%%%%%%%%%%%%%%%%%%%%%%%%%%%%%%%%%%%%%%%%%%%%
%%%%%%%%%%%%%%%%%%%%%%%%%%%%%%%%%%%%%%%%%%%%%%%%%%%%%%%%%%%%
\section*{Acknowledgments}

We are grateful to Gregory Korchemsky for illuminating correspondence,
and to Sven Moch for comments on the manuscript.
This work was
funded by the Deutsche Forschungsgemeinschaft (DFG, German Research
Foundation) Grant No.~460391856.
We acknowledge support from DESY (Hamburg, Germany), a member of the
Helmholtz Association HGF, and by the Deutsche
Forschungsgemeinschaft (DFG, German Research Foundation) under
Germany's Excellence Strategy – EXC 2121 ``Quantum Universe'' –
390833306.
A.\,B. is further supported by the Studienstiftung des Deutschen
Volkes.
This research was supported through the Maxwell computational
resources operated at Deutsches Elektronen-Synchrotron DESY, Hamburg,
Germany.

%%%%%%%%%%%%%%%%%%%%%%%%%%%%%%%%%%%%%%%%%%%%%%%%%%%%%%%%%%%%
%%%%%%%%%%%%%%%%%%%%%%%%%%%%%%%%%%%%%%%%%%%%%%%%%%%%%%%%%%%%
\arxor{}{
	\section*{Data availability}
	The data and code supporting the findings of this work are openly available as ancillary files associated with Ref.~\cite{Bargheer:2025hvd}. The ancillary files include the solution files and the code necessary to reproduce the results presented in this work.	
}

\ifarxiv

\appendix

%%%%%%%%%%%%%%%%%%%%%%%%%%%%%%%%%%%%%%%%%%%%%%%%%%%%%%%%%%%%
%%%%%%%%%%%%%%%%%%%%%%%%%%%%%%%%%%%%%%%%%%%%%%%%%%%%%%%%%%%%
\section{The Konishi Dimension}
\label{sec:konishiFull}

For reference, we record the complete five-loop scaling dimension of the
Konishi operator:
\begin{align}
	\Delta_{\mathcal{K}}
	&= 2 + 3 a - 3 a^2 + \frac{21}{4}a^3
	\nn\\ & \quad
	+ \brk*{
		-\frac{39}{4} + \frac{9}{4}\zeta_3
		- \frac{45}{8} \brk[s]*{1+\frac{12}{\Nc^2}} \zeta_5
	} a^4
	\nn\\ & \quad
	+ \Bigg(
		\frac{237}{16} + \frac{27}{4}\zeta_3
		-\frac{135}{16} \brk[s]*{1 - \frac{16}{\Nc^2}} \zeta_5
	\nn\\ & \quad\quad\quad
		-\frac{81}{16} \brk[s]*{1 + \frac{12}{\Nc^2}} \zeta_3^2
		+\frac{189}{32} \brk[s]*{5 + \frac{63}{\Nc^2}} \zeta_7
	\Bigg) a^5
	\nn\\ & \quad
	+ \order{a^6}
	\nn\\ &\approx
	2 + 3a -3a^2 + 5.25a^3
	- \brk*{12.878 + \frac{69.993}{\Nc^2}} a^4
	\nn\\ & \quad
	+ \brk*{36.64 + \frac{427.406}{\Nc^2}} a^5
	+ \order{a^6}
	\,.
	\label{eq:fullKonishiDimension}
\end{align}
The planar parts were found in~\cite{Anselmi:1996mq},
\cite{Eden:2000mv,Bianchi:2000hn},
\cite{Kotikov:2004er},
\cite{Fiamberti:2008sh,Velizhanin:2008jd,Bajnok:2008bm},
and~\cite{Bajnok:2009vm,Arutyunov:2010gb,Balog:2010xa} at one, two,
three, four, and five loops, respectively.
The four-loop $1/\Nc^2$ non-planar correction term was found in~\cite{Velizhanin:2009gv}.
The planar parts were re-derived in~\cite{Eden:2012fe}, which also
established the method employed in this paper.

It is known that the weak-coupling expansion of the planar theory in
the $\Nc\to\infty$ limit converges for $a<1/4$~\cite{Beisert:2006ez}.
The coefficients in the Konishi
dimension~\eqref{eq:fullKonishiDimension} indicate that the radius of
convergence will shrink at finite~$\Nc$, as expected in a general
Yang--Mills theory. A plot of the Konishi dimension for
different values of $\Nc$ is shown in
\figref{fig:konishiDimension}. Writing the Konishi dimension as
\begin{equation}
\Delta_{\mathcal{K}}=2 + \sum_{\ell=0}^{\infty} \gamma_\mathcal{K}^{(\ell)} a^\ell
\,,
\end{equation}
we can estimate the radius of convergence as
\begin{equation}
\lim_{\ell\to\infty}
\abs*{\frac{ \gamma_\mathcal{K}^{(\ell)}}{ \gamma_\mathcal{K}^{(\ell-1)}}}
\approx
\begin{cases}
1/0.25 & \Nc=\infty \,, \\
1/0.16 & \Nc=3 \,,
\end{cases}
\label{eq:RadiusEstimate}
\end{equation}
see \figref{fig:Radius}. The number $0.25$ for $\Nc=\infty$ is
expected, and the lower number $0.16$ confirms that the radius of
convergence shrinks at finite $\Nc$.

\begin{figure}
\centering
\includegraphics{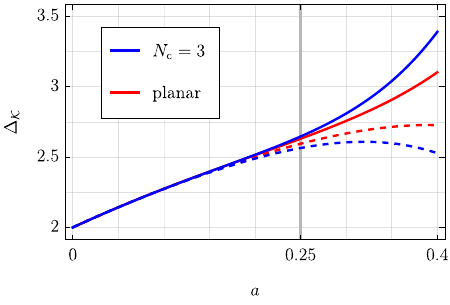}
\caption{The Konishi Dimension $\Delta_{\mathcal{K}}$ as a function of the
coupling~$a$ for $\Nc=\infty$ (red) and $\Nc=3$ (blue) at four
loops (dashed) and five loops (solid). The radius of convergence of the
planar theory is at $a=1/4$.}
\label{fig:konishiDimension}
\end{figure}
%

%%%%%%%%%%%%%%%%%%%%%%%%%%%%%%%%%%%%%%%%%%%%%%%%%%%%%%%%%%%%
%%%%%%%%%%%%%%%%%%%%%%%%%%%%%%%%%%%%%%%%%%%%%%%%%%%%%%%%%%%%
\section{Twistors}
\label{sec:twistors}

$\mathcal{N}=4$ SYM theory can be formulated in the super-twistor
space $\mathbb{CP}^{3|4}$, parameterized by homogeneous coordinates
$\mathcal{Z}=(Z^I, \chi^A)$, consisting of complex three-dimensional
projective coordinates $Z^I$ and four Grassmann degrees of freedom
$\chi^A$. They are related to points in chiral Minkowski
super-spacetime $(x^{\alpha\dot{\alpha}}, \Theta^{A\alpha})$ via the
incidence relations
\begin{equation}
	Z^I = (\lambda_\alpha, ix^{\dot\alpha\beta}\lambda_\beta) \quad\text{and}\quad \chi^A = \Theta^{A\beta}\lambda_\beta\,,
\end{equation}
with $\lambda_\alpha$ being homogeneous coordinates. The fields
in the theory are gathered in a superfield $\mathcal{A}(\mathcal{Z})$, that takes values in $(0, 1)$-forms and belongs to the adjoint
representation of $\grp{SU}(\Nc)$.
The twistor action takes the form~\cite{Boels:2006ir}
\begin{align} \label{eq:twistoraction}
	S(\mathcal{A}) =& \int_{\mathbb{CP}^{3|4}}\mathcal{D}^{3|4}\mathcal{Z}\wedge\tr\left(\frac{1}{2}\,\mathcal{A}\,\bar\partial\mathcal{A}-\frac{1}{3}\,\mathcal{A}^3\right)
	\nonumber \\
	&+\int d^4x\,d^8\Theta\,  L_\mathrm{int}(x, \Theta)\,,
\end{align}
where $\mathcal{D}^{3|4}\mathcal{Z}$ is the canonical holomorphic
volume on $\mathbb{CP}^{3|4}$. The first term is the self-dual part of
the theory, whereas the second term describes the non-local interaction
part
\begin{equation}
	L_\mathrm{int}(x, \Theta) = g_\mathrm{YM}^2\big(\log\det(\bar\partial-\mathcal{A})-\log\det\bar\partial\,\big)\,.
\end{equation}
\begin{figure}
\centering
\includegraphics{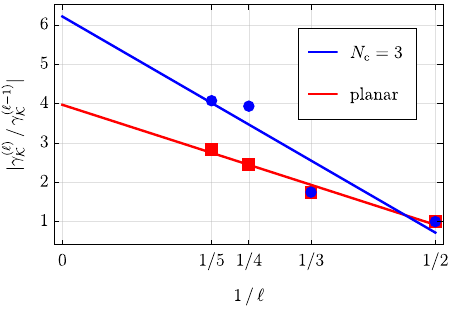}
\caption{Estimate of the convergence radius of the
Konishi dimension weak-coupling expansion, for $\Nc=\infty$ (squares, red) and for
\mbox{$\Nc=3$} (circles, blue). Shown are the ratios $\protect\abs{ \gamma_\mathcal{K}^{(\ell)}/ \gamma_\mathcal{K}^{(\ell-1)}}$ as
in~\eqref{eq:RadiusEstimate}, as a function of $1/\ell$, as well as
linear fits whose intersections with $1/\ell=0$ yield the numbers
in~\eqref{eq:RadiusEstimate}.}
\label{fig:Radius}
\end{figure}%
At this point, we can introduce the axial gauge, in which the cubic
term of the self-dual part of the action vanishes, simplifying the
following perturbative Feynman computations. In return, a fixed choice
of a reference twistor $Z_\star$ is introduced. While physical
observables remain independent of this choice, all intermediate
expressions will generally depend on it. Then, the propagator is
defined by the remaining quadratic term
\begin{equation}
	\langle \mathcal{A}^a(\mathcal{Z}_1)\mathcal{A}^b(\mathcal{Z}_2)\rangle = \bar\delta^{2|4}(\mathcal{Z}_1, \mathcal{Z}_2,\mathcal{Z_\star})\,\delta^{ab}\,,
\end{equation}
where $a, b$ denote the color indices of the fields, and $\bar\delta$
is a projective delta function. In order to describe the chiral,
half-BPS stress-tensor multiplet $\mathcal{O}(x, y, \theta)$, we
decompose the chiral superspace into two parts
\begin{equation}
	\Theta^{A\alpha} = W^{A}_a\theta^{a\alpha} + Y^A_{a'}\Psi^{a'\alpha}\,,
\end{equation}
with the half-BPS condition implying the independence of the stress-tensor multiplet from $\Psi^{a'\alpha}$. In supertwistor space, it is thus characterized by a subspace $\mathbb{CP}^{1|2}$, that can be parameterized by $(\lambda,\psi)$ in the following way
\begin{align}
	Z(\lambda)&=\lambda^1 Z_{1}+\lambda^2 Z_{2}=\lambda^\beta Z_{\beta}\,,\\
	\eta(\lambda,\psi)&=\lambda^\alpha \theta^a_\alpha W_{a}+\psi^{a'}Y_{a'}\,,
\end{align}
with $\psi^{a'} = \Psi^{a'\alpha} \lambda_\alpha$ and two choices of
$Z_\beta$ that define the twistor line corresponding to the point~$x$.
The stress-tensor multiplet can be expressed in twistor space as~\cite{Chicherin:2014uca}
\begin{equation}
	\mathcal{O}(x, y, \theta) = \int d^4\Psi\,L_\mathrm{int}(x, \Theta)\,.
\end{equation}
With these ingredients at hand, we can formulate the four-point
loop-integrand $G_4^{(\ell)}$ as follows:
\begin{equation}
	G_4^{(\ell)} = \eval*{\int d^4\theta_5\ldots d^4\theta_{4+\ell} \,\langle\mathcal{O}_1 \ldots \mathcal{O}_{4+\ell}\rangle_{ \mathrm{tree}}}_{\theta_1,\ldots,\theta_4\rightarrow0}\,,
\end{equation}
where the expectation value
is evaluated with respect to the twistor action~\eqref{eq:twistoraction} at tree level, effectively setting $\Lint=0$.
Moreover, we abbreviate
$\mathcal{O}(x_i, y_i, \theta_i)=\mathcal{O}_i$.
Applying the axial gauge allows for a straightforward derivation of Feynman rules for the above correlation function. The connection between these rules and the usual Minkowski spacetime coordinates and polarization vectors can be made explicit by adopting the following parametrization
\begin{align}
	(Z_{i,\beta})_{\alpha}^{\dot\alpha}
	&=\left(\epsilon_{\alpha\beta},x_{i,\beta}^{\dot\alpha}\right)\,, \nonumber\\
	W^B_{i,a}&=\left(\delta^b_a,\,0\right)\,, \nonumber \\
	Y^B_{i,a'}& =\left(y^b_{i,a'},\,\delta^{b'}_{a'}\right)\,.
\end{align}
With this choice, the resulting $4\times 4$ determinants (denoted by angle brackets) compute the basic Lorentz and R-charge invariants, with unit proportionality factors
\begin{align}
	\langle Z_{i,1}Z_{i,2} Z_{j,1}Z_{j,2}\rangle &\equiv \det (x_i-x_j)^{\dot\alpha \beta}=x_{ij}^2\,,\\
	\langle Y_{i,1}Y_{i,2} Y_{j,1}Y_{j,2}\rangle &\equiv \det (y_i-y_j)^{a b'}=y_{ij}^2\,.
\end{align}
While the twistor Feynman rules and their application was discussed in
the main text, here we elaborate further on the structure and explicit
expression of the $R$-factors that dress the Feynman graph vertices.
As explained, each operator $i$ that is
connected to more than two vertices $j_1\,\ldots,j_k$ is dressed by an
$R$-factor $R^i_{j_1\,\ldots,j_k}$, that exhibits a cyclic symmetry in
its lower indices. These factors can be split into $R$-factors of
valency three by using
\begin{equation}
	R^i_{j_1 j_2\dots j_k}
	=R^i_{j_1 j_2 j_3}R^i_{j_1 j_3 j_4}\dots R^i_{j_1j_{k-1}j_k}\,,
\label{eq:Rsplit}
\end{equation}
where each valency-three $R$-factor is given by
\begin{equation} \label{eq:R}
	R^i_{jkl} = -\frac{\delta^{0|2}\left(\langle ijk \rangle \psi_{il}+\langle ikl \rangle \psi_{ij}+\langle ilj \rangle \psi_{ik}\right)}{\langle ijk \rangle\langle ikl \rangle \langle ilj \rangle}\,.
\end{equation}
Here, we defined $\langle ijk \rangle = \epsilon_{\alpha\beta}\lambda_{ij}^\alpha \lambda_{ik}^\beta$. The parameters $\lambda_{ij}$ and $\psi_{ij}$ satisfy the following on shell conditions
\begin{align}
	\lambda_{ij}^\alpha &= \epsilon^{\alpha\beta}\frac{\langle Z_{i, \beta} Z_\star Z_{j, 1} Z_{j, 2}\rangle}{\langle Z_{i, 1} Z_{i, 2} Z_{j, 1} Z_{j, 2}\rangle}\,, \nonumber\\
	\psi_{ij}^{a'} &= \epsilon^{a'b'}\frac{\langle Y_{i, b'} E_{ij} Y_{j, 1} Y_{j, 2}\rangle}{\langle Y_{i, 1} Y_{i, 2} Y_{j, 1} Y_{j, 2}\rangle} \,,
	\label{eq:onshellcond}
\end{align}
where we used the definition
\begin{equation}
	E_{ij}^A=\lambda^\alpha_{ij}(\theta_i)^a_\alpha W_{i,a}^A +\lambda^\alpha_{ji}(\theta_j)^a_\alpha W_{j,a}^A\,.
\end{equation}
Each of the $R$-factors \eqref{eq:R} is completely antisymmetric in
the lower indices and of homogeneous degree two in the Grassmann
variables $(\theta_i)^a_\alpha$. Graphs that can
contribute to the loop-integrand $G_4^{(\ell)}$ yield expressions
that contain products of exactly $2\ell$ $R$-factors. In order to
contribute to the final result, such products must contain the
unique product $\theta_5^4\ldots\theta_{4+\ell}^4$ of all
non-zero Grassmann variables.

%%%%%%%%%%%%%%%%%%%%%%%%%%%%%%%%%%%%%%%%%%%%%%%%%%%%%%%%%%%%
%%%%%%%%%%%%%%%%%%%%%%%%%%%%%%%%%%%%%%%%%%%%%%%%%%%%%%%%%%%%
\section{Grassmann Contraction} \label{app:grassmann}

As explained in the main text, we are applying the iterative procedure
\arxor{\eqref{eq:grassmann}}{(16)} to numerically perform the Grassmann algebra
required to evaluate the twistor expression. The algorithm starts by
multiplying the first two $R$-factors to a purely quartic polynomial with
$\binom{20}{4}=4845$ terms corresponding to all ordered products of four Grassmann
variables. Then, this intermediate result is multiplied with the next
$R$-factor, arriving at a polynomial of a degree six. In this
manner, we successively multiply one
$R$-factor at a time, eventually arriving at the final unique product involving
all $20$ Grassmann numbers. At each intermediate step, the degree
${2a}$ polynomial is represented as a vector of $\binom{20}{2a}$
coefficients, and the multiplication
is performed by an order-three tensor $t_{a-1}$
of size
$\binom{20}{2a}\times\binom{20}{2a-2}\times\binom{20}{2}$, whose
non-zero entries are $\pm 1$. The number of non-zero elements of each
tensor can be computed by
\begin{equation}
	N_{a-1} = \binom{2a}{2} \times \binom{20}{2a}\,.
\end{equation}
The first factor counts the number of ways to decompose a
degree~$2a$ monomial into two monomials
of degrees 2 and $2a-2$,
while the second factor corresponds to the number of terms in a
degree~$2a$ polynomial (\ie the range of the last index of $t_{a-1}$).
As such, the tensors are very sparse, facilitating the computation. We
summarize the properties of the tensors in~\tabref{tab:tensors}.

\begin{table}[t]
	\centering
	{
		\setlength{\tabcolsep}{10pt}
		\begin{tabular}{rrr}
            \toprule
			$a$  & size & $N_a$  \\
            \midrule
			1    & $4845\times190\times190$ &  \num{29070}    \\
			2    & $38760\times4845\times190$ & \num{581400}   \\
			3    & $125970\times38760\times190$ &  \num{3527160}  \\
			4   & $184756\times125970\times190$ &  \num{8314020}   \\
			5   & $125970\times184756\times190$ &  \num{8314020}   \\
			6   & $38760\times125970\times190$ &  \num{3527160}   \\
			7   & $4845\times38760\times190$ &  \num{581400}    \\
			8   & $190\times4845\times190$ &  \num{29070}     \\
			9   & $1\times190\times190$ &  \num{190}       \\
			\bottomrule
		\end{tabular}
	}
    \caption{Properties of the order-three tensors $t_a$ appearing
    in the algorithm for the Grassmann contraction~\arxor{\eqref{eq:grassmann}}{(16)}.
    Listed are the respective sizes of the tensors, as well as the
    number of non-zero entries $N_a$.}
    \label{tab:tensors}
\end{table}

We can estimate the total computation time required to evaluate the
integrand numerically on a single kinematic point: We need to add the
contributions of ${\approx}\,87\,\mathrm{M}$ graphs (see \arxor{\tabref{tab:twistor}}{Table II}). For
each graph, $\sum_{a=1}^9N_a\approx{25\,\mathrm{M}}$
numerical multiplications must be performed, and equally many
additions. Assuming 8 FLOPS per cycle per core (using AVX instructions),
$3\,\mathrm{GHz}$ clock speed and $24$ cores, computing one data point
would need
\begin{equation}
T\approx\frac{2\times87\times25\times{10^{12}}}{24\times8\times3\times{10^9}/\mathrm{s}}
\approx 128 \, \text{Minutes}
\,.
\end{equation}
However, performing the computation with \mathematica on an AMD EPYC
74F3 24-core CPU, we find that the above estimate is significantly
over-optimistic, and that the actual runtime is about two orders of
magnitude bigger. The most likely reason is that our process is
I/O-bound, \ie the CPU cannot be fed with data fast enough to use its
full potential (the tensors do not fit in the L1 cache). Also,
\mathematica might not be optimized for this type of operation.

Since the numerical computation is highly parallelizable, we opt to
run the sequence of ten sparse tensor-vector-vector multiplications on
a GPU, which significantly accelerates the computation. We use the
\texttt{cuda.jit} decorator from the
Python package Numba \cite{numba}, which compiles the function
into a custom CUDA kernel
for efficient parallel execution. For each $a \in \set{2,\dots,10}$, the kernel
distributes the work of iterating over non-zero entries of the tensor
$t_{a-1}$
across CUDA threads, each of which independently multiplies a
single non-zero tensor element with the corresponding values in the
input vectors $S_{a-1}$ and $R_a$. The partial product is then
atomically accumulated into the output vector $S_a$ to avoid race
conditions.

To achieve high performance, we adopt a batched processing strategy in
which multiple tensor-vector-vector multiplications of the same tensor
but different $R$-factors are executed simultaneously. This batch
dimension allows us to better utilize GPU resources and increase
thread occupancy. For each multiplication, the kernel launch
configuration uses a specified number of threads per block and blocks per
grid, which control the parallel decomposition of the workload. Over
all contractions, we use a fixed number of threads per block, while the
number of blocks per grid is computed dynamically based on the number
of non-zero elements in each tensor and the batch size. The values of
the thread block size and batch dimension were tuned through
performance benchmarking to maximize throughput on the GPU
(see \figref{fig:performance}). With optimal setting of these
parameters, we find that the Grassmann contractions involved when
evaluating the twistor expression for one numerical kinematic point takes
roughly 17 hours on a single \textit{Nvidia A100} GPU.
The code for this procedure can be found in the ancillary file
\filename{07-numbaContract5Loop.py}.

\begin{figure}[tb]
	\centering
	\includegraphics[width=0.9\columnwidth]{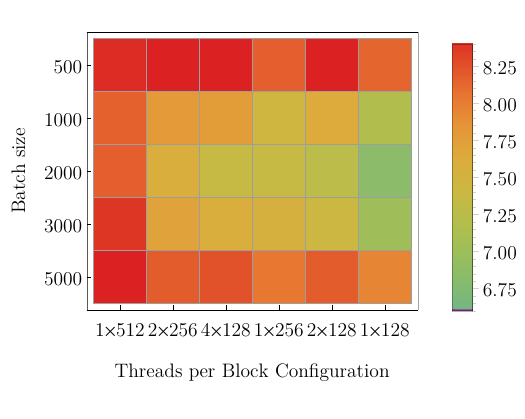}
	\caption{Heat map showing the time (in seconds) to perform 10,000
		five-loop contractions on an \textit{Nvidia A100} GPU. The y-axis
		represents the batch size, and the x-axis corresponds to the
		threads-per-block configuration.}
	\label{fig:performance}
\end{figure}

The numerical implementation of this contraction entails a significant
precision loss, such that not even \texttt{float64} numbers are
sufficient to ensure accurate results. For this reason, we opt to
evaluate both the twistor expression and the ansatz of the integrand
using \texttt{int32} integers within a finite field $\mathbb{F}_p$. To
ensure the intermediate values remain within the bounds of the
\texttt{int32} format, each tensor contraction must produce results
that do not exceed the $32$~bit integer limit. Considering that the
tensor multiplications involve at most 190 summands, we require that
$190\,p^2 < 2^{31}$, which implies that the largest prime we can
safely use is given by $p=3361$.

In the first instance, the particular solution (match between ansatz
and twistor expression) will only be valid over
the finite field $\mathbb{F}_{3361}$. However,
tuning the result by adding suitable combinations of the 208 Gram
identities, we can construct a particular
solution where the majority of the 930 $f$-graph coefficients is
zero, and all others attain small integer values (see
\figref{fig:histogram}). We expect this
to be the true solution over the real numbers. As a check, we compare
the particular solution against the twistor expression a few times
over different finite fields ($\mathbb{F}_{3347}$ and
$\mathbb{F}_{3359}$). We find perfect agreement, which proves the
validity of the solution beyond reasonable doubt.

\begin{figure}[tb]
	\centering
	\includegraphics[width=0.9\columnwidth]{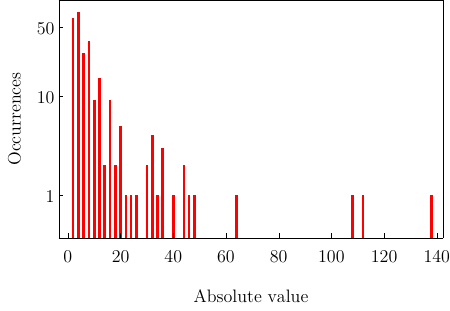}
	\caption{Histogram of absolute values in
		the particular genus-one five-loop integrand
		solution provided in the ancillary files, excluding $582$ zero coefficients.}
	\label{fig:histogram}
\end{figure}

%%%%%%%%%%%%%%%%%%%%%%%%%%%%%%%%%%%%%%%%%%%%%%%%%%%%%%%%%%%%
%%%%%%%%%%%%%%%%%%%%%%%%%%%%%%%%%%%%%%%%%%%%%%%%%%%%%%%%%%%%
\section{Non-Planar Master Integrals}
\label{app:masterInt}

In order to compute the sum of integrals~\arxor{\eqref{eq:masterInt}}{(23)} that
produces the Konishi anomalous dimension, we need to evaluate
two non-planar four-loop master integrals that have not been computed before.
Explicitly, the unknown integrals read (with $D=4-2\epsilon$)
\begin{align}
	\mathcal{I}_{3}
	&=
	\frac{1}{(\pi^2)^4}\int\frac{d^D x_6 d^D x_7 d^D x_8 d^D x_9}{\x{17}\x{18}\x{19}\x{57}\x{58}\x{59}\x{67}\x{68}\x{69}}
	\,,
	\\ \nonumber
	\mathcal{I}_4
	&=
	\frac{1}{(\pi^2)^4}\int\frac{d^D x_6 d^D x_7 d^D x_8 d^D x_9}{\x{16}\x{17}\x{18}\x{19}\x{56}\x{57}\x{58}\x{59}\x{67}\x{78}\x{89}\x{69}}
	\,,
\end{align}
see \figref{fig:two_graphs} for their diagrammatic representation.
Expanding the integrals to linear order in $\epsilon$ with a priori
undetermined coefficients yields (setting $\x{15}=1$)
\begin{align} \label{eq:masterIntAnsatz}
	\mathcal{I}_{3}
	&=
	\frac{a_3}{\epsilon} + b_3 + c_3 \epsilon +\order{\epsilon^2}
	\,,
	\nonumber \\
	\mathcal{I}_4
	&=
	\frac{a_4}{\epsilon} + b_4 + c_4 \epsilon +\order{\epsilon^2}
	\,.
\end{align}
To constrain the six unknown coefficients, we begin by deriving relations between
the $24$ master integrals by making use of the Gram
identities.
As discussed around \arxor{\eqref{eq:gram}}{(7)}, these identities manifest as $208$ linear
relations among the nine-point $f$-graphs.
By following the same steps that led to~\arxor{\eqref{eq:masterInt}}{(23)}, we can
transform these identities to relations among the $24$ two-point
four-loop master integrals.

\begin{figure}[t]
	\centering
	\begin{minipage}{0.235\textwidth}% half page width
		\centering
		\includegraphics[width=\textwidth]{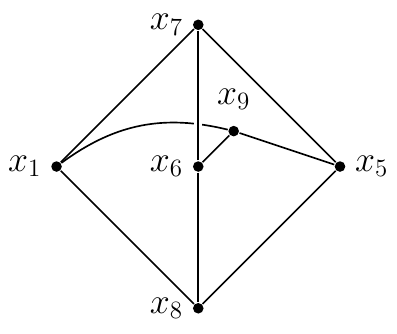}
		$\mathcal{I}_3$
	\end{minipage}
	\hfill
	\begin{minipage}{0.235\textwidth} % half page width
		\centering
		\includegraphics[width=\textwidth]{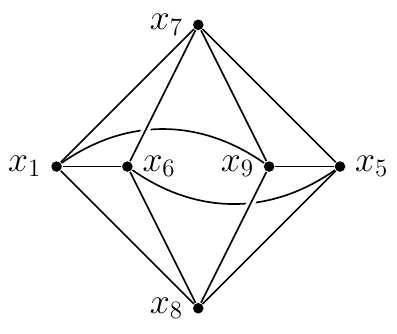}
		$\mathcal{I}_4$
	\end{minipage}
	\caption{Diagrammatic representation of the two
    non-planar four-loop master integrals that contribute to the
    Konishi anomalous dimension. The edges connecting vertices $i$ and $j$
    stand for factors $1/\x{ij}$.}
	\label{fig:two_graphs}
\end{figure}

First, we normalize the Gram identities among $f$-graphs in analogy to~\arxor{\eqref{eq:OPEfgraphs}}{(19)}. We then take the
OPE limit ($x_2\rightarrow x_1$,
$x_4\rightarrow x_3$), followed by $x_3\rightarrow\infty$, and finally
integrate over points $x_6,\dots,x_9$ in $D=4-2\epsilon$ dimensions.
Thus, we again arrive
at four-loop propagator-type integrals, that can be reduced to a
set of master integrals using \texttt{FIRE6}. In fact, they reduce to
the same set of $24$ master integrals that already appear in the
computation of $\gamma_\mathcal{K}^{(1,5)}$. Thus the Gram identities
turn into linear relations among the $24$ master integrals, with
$\epsilon$-dependent coefficients.

Plugging in the expressions ($\epsilon$ expansions) of the $20$ known
planar integrals~\cite{Baikov:2010hf,Lee:2011jt} and the two known
non-planar ones~\cite{Eden:2012fe}, we find constraints among the
coefficients in~\eqref{eq:masterIntAnsatz}. The constraints can be
brought to the following form:
\begin{align} \label{eq:masterIntConstraints}
	a_3 &= 0 \,,
	\quad b_3 = 36 \zeta_3^2 \,, \\
	a_4 &= -10 \zeta_5 \,,
	\quad b_4 = -\tfrac{5\pi^6}{189} + 2\zeta_3^2 + 50 \zeta_5 \,, \nonumber \\
	c_4 &= -\tfrac{32}{27}c_3 + \tfrac{25\pi^6}{189}
	+ \tfrac{67}{45}\pi^4 \zeta_3
	+ \tfrac{226}{3}\zeta_3^2
	+ 90\zeta_5
	- \tfrac{1143}{2}\zeta_7 \,. \nonumber
\end{align}
In total, four of the coefficients are fixed, while the remaining two
at linear order in $\epsilon$ are related by one equation. Notably,
$\mathcal{I}_3$ is finite in $\epsilon$. Let us mention, that this
solution is formulated in the $G$-scheme used also in~\cite{Baikov:2010hf,Lee:2011jt}.

To determine the final two coefficients~$c_3$ and~$c_4$, we evaluate~$\mathcal{I}_3$,
which is linearly reducible:
The integrations can be performed iteratively in a way that each
intermediate integral is a hyperlogarithm.
In addition, $\mathcal{I}_3$ involves comparatively few factors
$1/\x{ij}$, and therefore can be evaluated with \texttt{HyperInt}~\cite{Panzer:2014caa}.
We Fourier transform the integral to momentum
representation using
\begin{equation}
	\mathcal{F}\left[\frac{1}{x^2}\right] = \frac{1}{\pi^{2-\epsilon}} \int d^{4-2\epsilon}x\, \frac{e^{i p x}}{x^2} = 4^{1-\epsilon}\frac{\Gamma(1-\epsilon)}{(p^2)^{1-\epsilon}}\,,
\end{equation}
and evaluate the momentum integrals with non-integer propagator exponents.
In this way, we obtain the $\epsilon$-expansion of $\mathcal{I}_3$.
The result validates the values of $a_3$ and $b_3$ found
in~\eqref{eq:masterIntConstraints}, and determines the value of $c_3$. Finally,
$c_4$ is obtained from the last equation in~\eqref{eq:masterIntConstraints}:
\begin{align}
	c_3 &= \tfrac{6\pi^4}{5} \zeta_3 + 72 \zeta_3^2 - 567 \zeta_7 \,, \nonumber \\
	c_4 &= \tfrac{25\pi^6}{189} + \tfrac{\pi^4}{15}\zeta_3
	- 10 \zeta_3^2 + 90 \zeta_5 + \tfrac{201}{2}\zeta_7 \,.
\end{align}
We verify the solution of the integrals $\mathcal{I}_3$ and
$\mathcal{I}_4$ numerically using
\texttt{FIESTA}~\cite{Smirnov:2021rhf} up to four digits in precision,
and find perfect agreement with our exact coefficients.%
\footnote{We also cross-checked the results of these integrals with
the package \texttt{HyperlogProcedures} \cite{HP}, with consistent
outcomes. We thank Oliver Schnetz for correspondence.}

%%% Local Variables:
%%% mode: LaTeX
%%% TeX-master: "5LoopNP"
%%% End:

\fi

%%%%%%%%%%%%%%%%%%%%%%%%%%%%%%%%%%%%%%%%%%%%%%%%%%%%%%%%%%%%
%%%%%%%%%%%%%%%%%%%%%%%%%%%%%%%%%%%%%%%%%%%%%%%%%%%%%%%%%%%%
% \pdfbookmark[1]{\refname}{references}
\bibliography{references}
% \bibliography{myrefs}

\end{document}